\documentstyle[12pt]{article}
\title{One-loop effective potential in ${\cal N}=\frac{1}{2}$
generic chiral superfield model}

\author{O.D. Azorkina$^{1}$\footnote{azorkina@tspu.edu.ru},
A.T. Banin$^{2}$\footnote{atb@math.nsc.ru}, I.L.
Buchbinder$^{1}$\footnote{joseph@tspu.edu.ru}, N.G.
Pletnev$^{2}$\footnote{pletnev@math.nsc.ru}}

\date{\it $^{1}$Department of Theoretical Physics\\ Tomsk State Pedagogical
University\\ Tomsk 634041, Russia\\$^{2}$ Department of
Theoretical Physics\\ Institute of Mathematics, Novosibirsk, \\
630090, Russia}

\begin{document}
\maketitle

\begin{abstract}
We obtain the one-loop quantum corrections to the K\"ahlerian and
superpotentials in the generic chiral superfield model on
the nonanticommutative superspace. Unlike all previous works, we use a
method which does not require to rewrite a star-product of
superfields in terms of ordinary products. In the K\"ahlerian
potential sector the one-loop contributions are analogous to ones
in the undeformed theory while in the chiral potential sector the
quantum corrections contain a deformation parameter.

\end{abstract}

\thispagestyle{empty}
\newcommand{\be}{\begin{equation}}
\newcommand{\ee}{\end{equation}}
\newcommand{\bea}{\begin{eqnarray}}
\newcommand{\eea}{\end{eqnarray}}

Low-energy dynamics of open strings coupled to a constant
self-dual graviphoton field strength ${\cal C}_{mn}$ can be
efficiently described in the terms of field theories with deformed
supersymmetry \cite{1}. Remarkable class of such theories is
formulated in nonanticommutative  (NAC) ${\cal N}={1\over 2}$
superspace \cite{seib}. Nonanticommutativity means that the
multiplication of superfields is described in terms of so-called
star-product, which is a fermionic version of the Moyal product.
This allows us to work out the details of nonanticommutative
supersymmetric field theory using a calculus on conventional
${\cal N}=1$ superspace. The component actions of such theories
contain the additional couplings with a parameter ${\cal C}_{mn}$
what leads to reduction of supersymmetry.  Study of the various
aspects of supersymmetric theories on deformed  superspaces has
been carried out in a number of recent papers (see e.g. \cite{ren
wz}, \cite{bbpjhep}, \cite{ren sym} for $D=4$ models, \cite{D2}
for $D=2$ models and \cite{iv} for extended supersymmetric models
in harmonic superspace).

One of the most interesting problems in nonanticommutative
supersymmetric theories is a structure of divergences and
renormalizability. It is has been demonstrated that ${\cal N}=
{1\over 2}$ supersymmetric Wess-Zumino \cite{ren wz},
\cite{bbpjhep} and Yang-Mills \cite{ren sym} models are
renormalizable. To be more precise:  at the one-loop level
renormalizability is lost literally but it can be restored not
only at one loop but to all orders of perturbation theory after
adding to the classical action the specific terms depending on
deformation parameter \cite{ren wz}, \cite{bbpjhep}.  Divergences
are only logarithmic so the induced supersymmetric breaking is
soft. From the field-theoretical point of view this property looks
rather mysterious because such models contain the higher
dimensional operators and the naive power-counting arguments
formally show that the theories are not renormalizable at any
order of perturbation theory. However since all such models are
consistent only in the Euclidean space, where dotted and undotted
spinors are unrelated, the new vertices appearing with the
deformation parameter are not accompanied by their conjugates and
that provides renormalizability.

Non-linear sigma-models (bosonic or supersymmetric) attract much
attention due to the possibilities of an effective description of
infra-red physics. Generic chiral superfield model, which is a
generalization of the supersymmetric sigma-model, plays an
important role because its relation with the superstring theory as
a subsector of an effective theory of ${\cal N}=1$ supersymmetric
string vacua \cite{gsw}. Action of this model is written in terms
of the K\"ahlerian effective potential $K(\Phi, \bar\Phi)$ and
chiral $W(\Phi)$ and antichiral $\bar{W}(\bar\Phi)$
superpotentials. Such a model of chiral superfields can be treated
as an effective theory, suitable for description of phenomena at
energies much less than some fundamental scale. Besides, in some
cases, a vacuum structure of supersymmetric gauge theories is
defined in terms of  nonperturbatively induced superpotentials
$W(\Phi)$ and $\bar{W}(\bar\Phi)$.

In ordinary supersymmetric theories there exists the
non-renormalization theorem \cite{ns} which states that quantum
corrections to the (anti)holomorphic superpotential are very much
constrained and sometimes remains non-renormalizable, whereas the
K\"ahler potential in general gets quantum corrections. Full
one-loop corrections to the K\"ahler potential have been computed,
both in the Wess-Zumino model (see e.g. \cite{bky}) and in more
general renormalizable models \cite{renmod}, (for two-loop
corrections see the complete list of references in \cite{ni}).

The component structure of the generic ${\cal N}=\frac{1}{2}$
supersymmetric chiral model defined in the two-dimensional
\cite{D2} and in the four-dimensional \cite{abbp, D4} nonanticommutative
superspaces has been investigated in details recently. It has
been shown that these theories are described in a closed form by enough
simple deformations of the K\"ahler potential and (anti)holomorphic
superpotential.  Geometrically, this deformation can be interpreted as
a fuzziness in the target space controlled by the vacuum expectation
value of the auxiliary field. However the quantum properties of such
models have never been studied. In particular the problem of the
renormalizability and the problem constructing of the effective
action have not been addressed so far.

In this paper we study the quantum aspects of
generic chiral superfield model in NAC superspace. We compute the
divergent and leading finite one-loop corrections
to the K\"ahler potential and superpotential using a superfield loop
expansion (see e.g.  \cite{11}, \cite{12}) and the approximation of
slowly variating field.  Calculation techniques is formulated by a
way which preserves the local star-product structure of classical
action on the all stages of quantum analysis.  The divergence
structure of the model is clarified. We show that besides the
divergences preserving ${\cal N}=1$ supersymmetry there is a new
divergent structure explicitly containing the nonanticommutativity
parameter.

The classical action for the generic chiral superfield model
\begin{equation}\label{clact}
S=\int d^8z \,K_\star (\Phi,\bar\Phi)+\int d^6z \,W_\star(\Phi) +\int
d^6\bar{z} \,\bar{W}_\star (\bar\Phi)~,
\end{equation}
up to two derivatives is encoded in three functions of the chiral multiplet:
the
K\"ahler  potential $K_\star$ that is only required to be a real
function and (anti)chiral superpotentials $\bar{W}_\star, W_\star$
that are required to be (anti)holomorphic. The K\"ahler
potential and superpotentials are arbitrary functions of chiral
$\Phi(y, \theta)$ and antichiral $\bar\Phi (y,\theta,\bar\theta)$
superfields. The subscript $\star$ implies that all functions are
understood as expansions in the power series with $\star$-product.
The expansions of the K\"ahler potential and the superpotential are defined
as follows
\begin{equation}\label{expans}
K_{\star} = \sum_{n,
\bar{n}=0}^{\infty}K_{n\bar{n}}\frac{1}{n!\bar{n}!}\underbrace{\Phi\star\cdots\star\Phi}_{n}\star
\underbrace{\bar\Phi\star\cdots\star\bar\Phi}_{\bar{n}}|_s ~,\quad
W_{\star} = \sum_{n=0}W_{n}\frac{1}{n!}\underbrace{\Phi\star\cdots\star\Phi}_{n}~,
\end{equation}
and analogously for another superpotential $\bar{W}_{\star}$.
Further we will use notation and definition given in papers \cite{seib} and
\cite{11}, \cite{12}.

To calculate the one-loop correction we imply a
background-quantum splitting of the (anti)chiral fields $\Phi
\rightarrow \Phi + \phi$ and $\bar\Phi \rightarrow \bar\Phi +
\bar\phi$. In contrast to our previous paper \cite{bbpjhep} we
do not reduce the star products of superfields to their ordinary
products. Instead of that, we find the explicit operator $\hat{\cal
 H}_\star$ which is the second variation of the classical action
 (\ref{clact}). This operator is formulated completely in terms of
$\star$-product and defines the spectrum of quantum fluctuations
on a given background. It leads to the following form of the
one-loop effective action \footnote{Note that the functional
integral in the Euclidean space is defined as $\int {\cal D}\Phi
{\rm e}^{-S[\Phi]}$ and therefore the one-loop effective action
for a complex superfields is defined as $\Gamma =\mbox{Tr}\ln
S''[\Phi]$.}
\begin{equation}\label{ga}
\Gamma^{(1)}=\mbox{Tr}\ln_\star \hat{\cal H}_\star~.
\end{equation}
The operator $\hat{\cal H}_\star$ is a natural generalization of
the superspace type operators for the case of a deformed
superspace\footnote{The subscript $\star$ at the logarithm $\ln$
needs for consistency with the definition of the Green function
$G=-\frac{1}{H_\star}\star \delta(z-z')=\int^\infty_0 ds \,{\rm
e}_\star^{sH_\star}\star \delta(z-z')$. Then an arbitrary
variation $\delta H_\star$ of the operator $H_\star$ in formal
definition $\Gamma^{(1)}$ gives a correct expression $\delta
\Gamma^{(1)}=\mbox{Tr} H_\star^{-1}\star \delta
H_\star=\int^\infty_0 ds \,\mbox{Tr} {\rm e}_\star^{sH_\star}\star
\,\delta H_\star=\int^\infty_0 \frac{ds}{s} \mbox{Tr}( \delta {\rm
e}_\star^{sH_\star})=\delta \mbox{Tr}\ln_\star H_\star$. }.
Analogously to Ref. \cite{0310144} we call a functional of $D_A$
and $\Phi, \bar\Phi$ a star-local polynomial functional if it
includes an integral over superspace of a finite sum of monomials
so that every monomial is given in terms of star products of a
finite number of $D_A$ and $\Phi, \bar\Phi$ taken in the same
superspace point. We will see that there are two type of
contributions to the effective action. Leading contributions are
star-local. However there can be some kind of non-local
contributions, which are similar to the contributions from
non-planar diagrams.  But for the effective potential calculation
in the approximation of slowly varying background fields such
non-local contributions not enter into the game since they lead to
higher derivative operators. We mainly focus on the approximation
of the constant background
\begin{equation}\label{const}
D_{\alpha, (\alpha\dot\alpha)}\Phi = 0,\quad \bar{D}_{\dot\alpha,
(\alpha\dot\alpha)}\bar{\Phi}=0,
\end{equation}
where $\bar{D}_{\dot\alpha}=\frac{\partial}{\partial
\bar\theta^{\dot\alpha}}, \quad D_\alpha=\frac{\partial}{\partial
\theta^{\alpha}}+ i\bar\theta^{\dot\alpha}\frac{\partial}{\partial
y^{\alpha\dot\alpha}}$.

Firstly we emphasize the basic
properties of $\star$-product used in the further calculation.
Because for the mixed powers in the power expansion of the K\"ahler potential
we have $\Phi \star \bar\Phi \neq \bar\Phi \star \Phi$, we consider
the products of superfields to be always fully symmetrized
\cite{D4}
\begin{equation}\label{cycl} \Phi^{1} \star  ...\star
 \Phi^{n}\star \bar\Phi^{1} \star ...\star \bar\Phi^{m}
|_{s}=\frac{1}{n!m!}(\Phi^{1} \star  ...  \star \Phi^{n}\star
\bar\Phi^{1} \star  ...\star \bar\Phi^{m}+ Perm.)
\end{equation}
Then, using the cyclic property  $\int \Phi_1 \star \Phi_2 \star
... \Phi_n=\int \Phi_n \star \Phi_1 \star ... \star \Phi_{n-1}$
and rules
$$
\frac{\delta \Phi(z)}{\delta
\Phi(z')}=-\frac{1}{4}\bar{D}^2 \delta^8(z-z'), \quad \frac{\delta
\bar\Phi(z)}{\delta \bar\Phi(z')}=-\frac{1}{4}{D}^2
\delta^8(z-z')~,
$$
where $D^2=D^\alpha D_\alpha,
\bar{D}^2=\bar{D}^{\dot\alpha}\bar{D}_{\dot\alpha}$, one obtains
the equations of motion for the model under consideration
\begin{equation}
-\frac{1}{4}\bar{D}^2 K_1 + W_1=0, \quad -\frac{1}{4}{D}^2
K_{\bar{1}} + \bar{W}_{\bar{1}}=0~,
\end{equation}
where $K_1=\frac{\partial K_\star(\Phi,\bar\Phi)}{\partial \Phi}$
etc.

According to (\ref{ga}) we have to calculate a second functional
derivatives of the classical action (\ref{clact}). Part of the
action dependent on the K\"ahler potential leads to three types of
the derivatives: mixed derivative $\delta^2 S/\delta \Phi \delta
\bar\Phi$ and two non-mixed derivatives $\delta^2 S/\delta
\Phi^2$, $\delta^2 S/\delta \bar\Phi^2$. These derivatives contain
both right and left star multiplications. It is easy to check that
non-mixed second functional derivatives are equal to zero.
Really, the power expansion of the K\"ahler potential is
defined by (\ref{expans}) where all powers are given according to
the rule (\ref{cycl}). Then
second non-mixed derivative leads to the expression
\begin{equation}\label{k11}
\frac{\delta^2 }{\delta \Phi(z')\delta \Phi(z)}\int d^8z \, K_\star=
\sum_{n\bar{n}}K_{n\bar{n}}\frac{1}{(n-1)!\bar{n}!}\times
\end{equation}
$$ \underbrace{(-\frac{1}{4}\bar{D}^2)\Phi \star \cdots\star
(-\frac{1}{4}\bar{D}^2) \delta^8(z-z')\star \cdots\star
\Phi}_{n-2}\star \bar\Phi\star \cdots \star \bar\Phi|_{s}=0~, $$
where we also suppose a sum over all possible permutations of
$\Phi$ and $(-\frac{1}{4}\bar{D}^2) \delta^8(z-z')$ according to the
definitions (\ref{expans}, \ref{cycl}). The same property is true for
another non-mixed variation $\frac{\delta^2}{\delta
\bar\Phi(z)\delta \bar\Phi(z')}\int d^8z K_\star =0$. The mixed
derivative has a quite different structure
\begin{equation}\label{k1bar1}
\frac{\delta^2 }{\delta \bar\Phi(z')\delta \Phi(z)}\int d^8z\,
K_\star= \sum_{n\bar{n}}K_{n
\bar{n}}\frac{1}{(n-1)!\bar{n}!}\times
\end{equation}
$$
(-\frac{1}{4}\bar{D}^2)\underbrace{\Phi \star \cdots \star
\Phi}_{n-1} \star \underbrace{\bar\Phi \star \cdots
\star(-\frac{1}{4} {D}^2) \delta^8(z-z')\star \cdots\star
\bar\Phi}_{\bar{n}-1}|_{s} =K_{1\bar{1}}\star
\frac{1}{16}\bar{D}^2{D}^2 \delta^8(z-z')~.
$$
This is a consequence of
the following property for the star-product \cite{abbp}
\begin{equation}
\begin{array}{c}
f_1(\theta) \star \ldots\star  f_{n}(\theta)\star
\delta(\theta-\theta')\star g_1(\theta) \star \ldots \star
g_m(\theta) \\ = \int d^2 \pi\,  f_1(\theta+{\cal C}\pi)\star
\ldots \star f_n(\theta+{\cal C}\pi) \star g_1(\theta-{\cal
C}\pi)\star \ldots \star g_m(\theta-{\cal C}\pi){\rm
e}^{(\theta-\theta') \pi}~.
\end{array}
\end{equation}
Therefore using the approximation (\ref{const}) ($f(\theta+{\cal
C}\pi)=f(\theta)+{\cal C}\pi Df(\theta)+...=f(\theta)$) we obtain
\begin{equation}
\begin{array}{c}
 f_1(\theta) \star \ldots\star  f_{n}(\theta)\star \delta(\theta-\theta')\star
g_1(\theta) \star \ldots \star g_m(\theta) \\ =f_1(\theta) \star
\ldots\star f_{n}(\theta)\star g_1(\theta) \star \ldots \star
g_m(\theta) \star \delta(\theta-\theta')
\end{array}
\end{equation}
The results (\ref{k11}, \ref{k1bar1}) for the second functional
derivatives and the totally symmetrical form of the expansion
(\ref{expans}) lead to simplification of the calculation procedure.

For the parts of the action dependent on the chiral superpotential we
have
\begin{equation}\label{w2}
\frac{\delta^2 }{\delta \Phi(z')\delta \Phi(z)}\int d^6z\,
W_\star= \sum_{n}W_{n}\frac{1}{(n-1)!}\times
\end{equation}
$$ \underbrace{\Phi \star \cdots\star (-\frac{1}{4}\bar{D}^2)
\delta^8(z-z')\star \cdots\star \Phi}_{n-2} |_s
=W_2\star(-\frac{1}{4}\bar{D}^2) \delta^8(z-z')~, $$ and
analogously for antichiral superpotential.

The one-loop correction to the effective
potential is written as follows
\begin{equation}\label{dzita}
\Gamma^{(1)}=\mbox{Tr}\ln_\star \hat{\cal H}_\star=\int d^8z
\ln_\star \hat{\cal H}_\star \star \delta^8(z-z')|_{z=z'}
\end{equation}
where the operator of second functional derivatives has, according
to (\ref{k11}, \ref{k1bar1}) and (\ref{w2}), the form
\begin{equation}\label{hath} \hat{\cal H}_\star = \pmatrix{
K_{1\bar{1}}\frac{1}{16}{D}^2\bar{D}^2&\bar{W}_{\bar{2}}(-\frac{1}{4}{D}^2)\cr
W_{2}(-\frac{1}{4}\bar{D}^2)&K_{\bar{1}1}\frac{1}{16}\bar{D}^2{D}^2}~.
\end{equation}
It should be especially noted that because we deal with
nonanticommutative superspace, all functions are understood as
power expansions containing the star-products of superfields. In
principle, one can extract for nonanticommutative theories both
star-local and star non-local contributions but we focus only on
the star-local approximation.

Using matrix operator (\ref{hath}) we calculate the leading (K\"ahler
potential) and next-to-leading (chiral potential) contributions.
It means that
\begin{equation}\label{KW}
\Gamma^{(1)}=\Gamma^{(1)}_{K} + \Gamma^{(1)}_{W}~.
\end{equation}
To find $\Gamma^{(1)}_{K}$ it is sufficient to consider in the
operator (\ref{hath}) the constant background superfields
$\bar{W}_{\bar{2}}$ and $W_{2}$. For getting the
$\Gamma^{(1)}_{W}$ we have to treat these superfields as slowly
varying and take into account the leading terms in their spinor
derivatives. It is obviously that the contribution to the
antichiral potential $\Gamma_{\bar{W}}^{(1)}$ will be equal to
zero (see e.g. consideration in \cite{bbpjhep}).

Let us begin with $\Gamma^{(1)}_{K}$. In this case one can note
that the diagonal and off-diagonal blocks of the matrix operator
$\hat{\cal H}_\star$ are commute between each other and,
therefore, the logarithm of the matrix can be splitted off into
two parts. This fact allows us to rewrite such a contribution in the
effective action as
\begin{equation}
\Gamma^{(1)}_{K} =\mbox{Tr}\ln_\star \pmatrix
{K_{1\bar{1}}\frac{1}{16}D^2\bar{D}^2&0\cr
0&K_{\bar{1}1}\frac{1}{16}\bar{D}^2D^2}
\end{equation}
$$
+ \mbox{Tr}\ln_\star
\left(1+\pmatrix{0&\frac{1}{K_{1\bar{1}}}\star\bar{W}_{\bar{2}}(-\frac{D^2}{4\Box})\cr\frac{1}{K_{\bar{1}1}}\star
W_{2}(-\frac{\bar{D}^2}{4\Box})&0 }\right)~.
$$

After calculations the matrix trace and using
the projector property we obtain
\begin{equation}
\Gamma^{(1)}_{K}=\mbox{Tr}\ln_\star(K_{1\bar{1}})\frac{1}{16}\frac{D^2\bar{D}^2}{\Box}+\frac{1}{2}
\mbox{Tr}\ln_\star\left(1-\frac{1}{K_{1\bar{1}}}\star
\bar{W}_{\bar{2}}\star \frac{1}{K_{\bar{1}1}}\star
W_{2}\frac{1}{\Box}\right)\frac{1}{16}\frac{D^2\bar{D}^2}{\Box}
+c.c.
\end{equation}
Next point is to analyze the structure of divergence
and renormalization properties. The first term in
the dimensional regularization scheme is equal to zero. The second
term (along with the complex conjugated) gives us
\begin{equation}
\begin{array}{l}
\displaystyle \Gamma^{(1)}_{K}=\mu^{4-d}\int\, \frac{d^d p
}{(2\pi)^d p^2}\,\ln_{\star}\left(1+\frac{1}{K_{1\bar{1}}}\star
\bar{W}_{\bar{2}}\star \frac{1}{K_{\bar{1}1}}\star
W_{2}\frac{1}{p^2}\right) \\ \displaystyle=\frac{1}{(4\pi)^2}\cdot
(4\pi \mu^2)^{2-d/2}\frac{d/2}{\Gamma(d/2+1)}\int dp^2
(p^2)^{d/2-2}\ln_{\star}\left(1+\frac{1}{K_{1\bar{1}}}\star
\bar{W}_{\bar{2}}\star \frac{1}{K_{\bar{1}1}}\star
W_{2}\frac{1}{p^2}\right)~,
\end{array}
\end{equation}
where $\mu$ is a regularization parameter and $d$ is the
space-time dimension. Putting $d=4-\epsilon$ we have
\begin{equation}
\Gamma^{(1)}_{K} =\frac{1}{2(4\pi)^2}\int d^8z\,
\frac{1}{K_{1\bar{1}}}\star \bar{W}_{2}\star
\frac{1}{K_{\bar{1}1}}\star
W_{2}\star\left(\frac{1}{4\pi\mu^2}\frac{1}{K_{\bar{1}1}}\star
\bar{W}_{\bar{2}}\star \frac{1}{K_{1\bar{1}}}\star
W_{2}\right)^{-\frac{\epsilon}{2}} \Gamma(\frac{\epsilon}{2})~.
\end{equation}
Using the scheme of minimal subtractions one gets the divergent
\begin{equation}
\Gamma^{(1)}_{K\; div}=\frac{1}{16\pi^2\epsilon}\int d^8z
\,\frac{1}{K_{1\bar{1}}}\star \bar{W}_{\bar{2}}\star
\frac{1}{K_{\bar{1}1}}\star W_{2} ~,
\end{equation}
and finite part
\begin{equation}
\Gamma^{(1)}_{K\; fin}=-\frac{1}{32\pi^2}\int d^8z
\frac{1}{K_{1\bar{1}}}\star \bar{W}_{\bar{2}}\star
\frac{1}{K_{\bar{1}1}}\star W_{2}\star\left(\ln_{\star}
(\frac{1}{K_{1\bar{1}}} \star \bar{W}_{\bar{2}}\star
\frac{1}{K_{1\bar{1}}}\star W_{2}\frac{1}{\mu^2}) +\gamma\right)~,
\end{equation}
of $\Gamma^{(1)}_{K}$. Here $\gamma$ is the Euler constant.
When the deformation parameter ${\cal C}_{mn}=0$,
the obtained results coincides with ones for
undeformed theory \cite{bky}, \cite{renmod}, \cite{bcp}. We point out
that in the K\"ahlerian effective potential sector the whole dependence on
nonanticommutativity is stipulated only by star-product.

Let us analyze now a structure of the next-to-leading contribution to the
effective action $\Gamma^{(1)}_{W}$. For this purpose we take up a
simple form for $K_{1\bar{1}} = 1+ {\cal O}(\bar\Phi)$ and
$\bar{W}_{\bar{2}}=\bar{m}+{\cal O}(\bar\Phi)$. The arguments for
such a choice of approximation are quite natural
\cite{ns}, \cite{zan}: contribution to the effective superpotential
can not dependent on the coefficients of the antichiral
superpotential. Indeed we can promote each of these
coefficients to an antichiral superfield field, whose vev then gives the
coupling constants. Holomorphy tells us that these fields can not
appear in an integral over chiral superspace. Since we are
interested in computing the effective chiral superpotential, we can
consider the antichiral superpotential to be simplest what leads to
$\bar{W}_{\bar{2}}=\bar{m}$.

According to the procedure described in \cite{bbpjhep}, the
quantity $\Gamma^{(1)}_{W}$ can be found from (\ref{hath}) in the
following form
\begin{equation}
\Gamma^{(1)}_{W}=\frac{1}{2}\mbox{Tr}\ln_\star\left(1+\pmatrix{-\bar{m}
\frac{D^2\bar{D}^2}{16\Box^2}W_2 &0 \cr
-\frac{\bar{D}^2}{4\Box}W_2 &
0}\right)=\frac{1}{2}\mbox{Tr}\frac{D^2\bar{D}^2}{16\Box}\ln_\star(1-\frac{\bar{m}}{\Box}
W_2 )~.
\end{equation}
Further consideration is done analogously to one presented in
\cite{bbpjhep}. After writing the integral over whole
superspace as $\int d^8z =\int d^6 z \bar{D}^2$ we obtain the
expression for the one-loop correction to the chiral superpotential
in the form
\begin{equation}\label{w} \Gamma^{(1)}_{W} =\frac{1}{2}\int d^6z
\,\ln_\star\left(1 -\frac{\bar{m}}{\Box} W_{2}\right)\star
(-\frac{1}{4}\bar{D}^2) \delta^8(z-z')|_{z=z'}~.
\end{equation}
Analogously to \cite{11}, \cite{zan} such a form  can be also obtained after integrating out the
antichiral field from the action
\begin{equation}
S(\Phi,\bar\Phi)=\int d^8z \,\bar\Phi \Phi +\int d^6z \,W_\star(\Phi)+
\int d^6\bar{z}\, \frac{\bar{m}}{2}\bar\Phi^2~.
\end{equation}
For this purpose ones rewrite the linear and quadratic over
$\bar{\Phi}$ part of the above action as follows
\begin{equation} \int
d^8z \left( \frac{\bar{m}}{2}(\bar\Phi
+\frac{1}{\bar{m}}D^2\Phi)\frac{1}{\Box}\bar{D}^2 (\bar\Phi
+\frac{1}{\bar{m}}D^2\Phi) -\frac{1}{2\bar{m}}D^2\Phi
\frac{1}{\Box}\bar{D}^2 D^2\Phi\right)~.
\end{equation}
Now the antichiral superfield can be integrated out in the
functional integral. Replacing in the last term $\int
d^2\bar\theta$ with $\bar{D}^2$ we obtain the action
\begin{equation} S(\Phi)= \int d^6z\,
\left(-\frac{1}{2\bar{m}}\Phi \Box \Phi +W_\star(\Phi) \right)~.
\end{equation} This action leads to the one-loop effective action
in the form (\ref{w}).

Note that in the undeformed model the expression (\ref{w}) is equal to
zero as it should be in accordance with \cite{ns}. In the case
under consideration the non-zero result is stipulated by
nonanticommutativity.  After enough simple calculations within
dimensional regularization one gets
\begin{equation} \Gamma^{(1)}_{W}\label{Gamma_W}
=\frac{\bar{m}^2}{64\pi^2} \int d^6z\, W_{2} \star W_{2} \star
(\frac{\bar{m} W_{2}}{4\pi\mu^2})^{-\frac{\epsilon}{2}}
\Gamma(\frac{\epsilon}{2})\star \delta^2(\theta-\theta')|~.
\end{equation}
To clarify a structure of this result we transform the expression
(\ref{Gamma_W}) in the form without $\star$-product and keep only
the leading term in deformation parameter. One can obtain
\begin{equation} \Gamma^{(1)}_{W}
=-\frac{\bar{m}^2}{64\pi^2} \int d^6z\,\frac{1}{2}{\cal C}^2 W_2 Q^2
W_2 (\frac{\bar{m}W_2}{4\pi \mu^2})^{-\frac{\epsilon}{2}}
\Gamma({\frac{\epsilon}{2}})~.
\end{equation}
This expression has the divergent part in the form
\begin{equation}\label{Wd}
\Gamma^{(1)}_{W\; div}
=-\frac{\bar{m}^2}{64\pi^2\epsilon} {\cal C}^2 \int d^6z \, W_{2}Q^2
W_{2}~,
\end{equation}
and a finite part. We point out that the divergent part (\ref{Wd})
explicitly contains the deformation parameter which can not be
absorbed into $\star$-product.

In the finite part we will try to restore the $\star$-product
under the integral over chiral subspace.  It leads to
\begin{equation}\label{Wf}
\Gamma^{(1)}_{W\;
fin} =\frac{\bar{m}^2}{128\pi^2}{\cal C}^2 \int d^6z \, W_{2} Q^2 W_{2}
\star \ln_{\star}(\frac{\bar{m} W_{2}}{\mu^2}) ~,
\end{equation}
where $Q=i\partial/\partial\theta$. It is clear that in the
expression (\ref{Wf}) the deformation parameter can not be
absorbed into initial $\star$-product. We already pointed out the
analogous situation in the expression (\ref{Wd}).

The relations (\ref{Wd}) and (\ref{Wf})
define the one-loop chiral effective potential in the model under
consideration.  In the case of Wess-Zumino nonanticommutative model,
the divergent term $\Gamma^{(1)}_{W\; div}$ takes the form $\Phi Q^2
\Phi$ which was recently found  (see e.g \cite{ren wz}, \cite{bbpjhep})
.

To conclude, we have presented calculations of the one-loop
effective potential for the nonanticommutative generic chiral
superfield model. We used the approximation of slowly variating
superfields and developed the method which allows to carry out the
calculations procedure without explicit rewriting the
$\star$-product in terms of ordinary products. The divergence
structure of the model is analyzed, we show that besides the
divergences analogous to ones for undeformed model, there is a new
divergent structure containing the nonanticommutativity parameter
and destructing the star-product structure of the model on a quantum
level. As a result, the divergent and finite one-loop K\"ahlerian
and chiral effective potentials are found in the explicit forms.

\section*{Acknowledgment}
I.L.B is grateful to Max Planck Institute of Physics (Munich)
where the paper has been finalized and D. L\"ust for kind hospitality.
The work was supported in part by RFBR grant, project No
03-02-16193. The work of O.D.A and I.L.B was also partially supported by
INTAS grant, INTAS-03-51-6346 and grant for
LRSS, project No 1252.2003.2. I.L.B is grateful to joint RFBR-DFG grant, project No
02-02-04002 and  DFG grant, project No 436 RUS 113/669 for partial support.
The work of N.G.P was supported in
part by RFBR grant, project No 05-02-16211.

\end{document}